# Initial Characterization of Stellar Photometry of Roman Images from the OpenUniverse Simulations

L. Aldoroty[1,4,5], D. Scolnic[1], A. Kannawadi[1], R. A. Knop[2], B. M. Rose[3], R. Hounsell[4,5], M. A. Troxel[1] The Roman Supernova Project Infrastructure Team
[1] Department of Physics, Duke University, Durham, NC 27708, USA; laldoroty@umbc.edu
[2] Lawrence Berkeley National Laboratory, 1 Cyclotron Road, MS 50B-4206, Berkeley, CA 94720, USA
[3] Department of Physics and Astronomy, Baylor University, One Bear Place #97316, Waco, TX 76798-7316, USA
[4] University of Maryland Baltimore County, 1000 Hilltop Cir, Baltimore, MD 21250, USA
[5] NASA GSFC, 8800 Greenbelt Rd, Greenbelt, MD 20771, USA

## Abstract

NASA's Nancy Grace Roman Space Telescope (Roman) will provide an opportunity to study dark energy with unprecedented precision and accuracy using several techniques, including measurements of high-$z$ Type Ia Supernovae (SNe Ia; $z \lesssim 3.0$) via the High-Latitude Time Domain Survey (HLTDS). In this work, we do an initial "benchmark" characterization of the photometric repeatability of stellar fluxes, which must be below 1% when sky noise is subdominant in order to enable a number of calibration requirements. Achieving this level of flux precision requires attention to Roman's highly structured, spatially varying, undersampled point-spread function (PSF). In this work, we build a library of effective PSFs (ePSFs) compatible with the OpenUniverse HLTDS simulations. Using our library of ePSFs, we recover stellar flux to between 0.6% and 1.2% photometric precision, finding that redder bands perform better by this metric. We also find that flux recovery is improved by up to 20% when a chip (sensor chip assembly, SCA) is divided into eight sub-SCAs in order to account for the spatial variation of the PSF. With our optimized algorithm, we measure nonlinearity due to photometry (magnitude dependence) of $|s_{\rm NL}| < 1.93 \times 10^{-3}$ dex$^{-1}$, which is still larger than stated requirements for detector effects and indicates that further work is necessary. We also measure the dependence of photometric residuals on stellar color and find the largest possible dependence in R062, implying a color-dependent PSF model may be needed. Finally, we characterize the detection efficiency function of each OpenUniverse Roman filter, which will inform future studies.

## 1. Introduction

The Nancy Grace Roman Space Telescope (Roman) is NASA's next planned flagship mission (D. Spergel et al. 2013, 2015; R. Akeson et al. 2019), scheduled to launch fall of 2026. It features the Wide Field Instrument (WFI), which will collect both imaging and slitless spectroscopy. The WFI will be sensitive to both optical and near-infrared wavelengths (0.48–2.3 $\mu$m), with a relatively large 0.281 deg$^2$ field of view.

One of Roman's primary science goals is to study dark energy with Type Ia Supernovae (SNe Ia). In accordance with the recommendations outlined in A. Albrecht et al. (2006), Roman is a Stage IV cosmology experiment and will enable multiple analysis techniques, as well as increase the dark energy figure of merit by 1 order of magnitude (D. Spergel et al. 2015). It will achieve these objectives partially through the High-Latitude Time Domain Survey (HLTDS), which is a core community survey designed to discover approximately $10^4$ SNe Ia out to $z \lesssim 3$ (R. Hounsell et al. 2018; B. M. Rose et al. 2021; B. Rose et al. 2025; R. Kessler et al. 2025; D. Rubin et al. 2025). Constraining power on cosmological parameters is due to the comparison of low-$z$ SNe Ia (at $z \approx 0.25$, $m_{{\rm peak},Y106} \approx 22$) up to high-$z$ SNe Ia (at $z \approx 2.75$, $m_{{\rm peak},Y106} \approx 28$). Therefore, any systematic uncertainties in photometry, particularly those with a dependence on brightness, is critical for cosmological measurements. Furthermore, for Roman, absolute calibration will rely on measuring bright spectrophotometric standards ($m_{Y106} \approx 18$) to set zero points for determining apparent magnitudes of other objects, including very faint SNe Ia. Thus, because the magnitude range of measurements for a Roman SN survey is so large, understanding the Roman photometric precision and accuracy is especially important.

A correct understanding and utilization of the Roman point-spread function (PSF) is a central facet of accurate flux recovery. However, the Roman PSF is not a straightforward element of data analysis; its PSF is expected to vary significantly across the focal plane. Because of Roman's wide field of view, the PSF is both spatially varying and undersampled. Some chromatic dependence is also expected (F. Berlfein et al. 2025). Due to the similarities between Roman and the Hubble Space Telescope (HST) PSFs, analysis of HST imaging provides a blueprint for Roman analyses. For example, like Roman's WFI, images from HST's Wide Field Camera (WFC) are undersampled. To address this challenge, J. Anderson & I. R. King (2000) developed a method to generate the "effective PSF" (ePSF), which includes a convolution of the instrumental PSF with the pixel sensitivity profile. The ePSF is an empirical model built from stars in a given image that describes the amount of flux from a star that falls into adjacent pixels as a function of that pixel's position with respect to the star's location. This method requires dithered observations in order to accurately determine both the star's position and the PSF. J. Anderson & I. R. King (2000) also suggested that in order to handle spatial variation in the PSF across a chip, the chip should be divided into a $3 \times 3$ grid and a PSF should be generated separately for each of the nine areas on the chip delineated by the grid. Then, linear interpolation between four of these PSFs to the specific location of the star provides a sufficient PSF. This work is followed up in J. Anderson (2016), where WFC3/IR PSFs were generated for community use. To account for spatial PSF variation, they divide the detector into a $3 \times 3$ grid and release nine PSFs per filter. The ePSF algorithm has been implemented in `photutils` (L. Bradley et al. 2024) for community use.

In order to evaluate the effectiveness of these techniques for Roman, we process images from the Roman OpenUniverse et al. (2025; hereafter OpenUniverse2024) simulations. End-to-end simulations like the OpenUniverse2024 images are critical for large mission preparation. They are designed to be realistic and include astrophysical, instrumental, and observing strategy effects. The OpenUniverse2024 images feature surveys from both Roman and LSST with overlapping viewing regions; however, in this work, we focus solely on the Roman HLTDS images.

In this work, we develop a library of ePSFs for the OpenUniverse2024 simulations. In Section 2.1, we describe the OpenUniverse2024 simulations. Section 2.2 describes the methodology used to generate our ePSF library, and Section 2.3 describes the photometric methods used to evaluate the performance of the ePSF library. We describe our source detection technique in Section 2.4. Section 3.1 discusses the results in terms of flux precision, and Section 3.4 describes detection efficiency results. Finally, we outline our conclusions and make recommendations for future work in Section 4.[6] This work is a contribution from the Roman Supernova Project Infrastructure Team (SNPIT), and can be used by the Roman community.

## 2. Methods

### 2.1. OpenUniverse

In this work, we use the OpenUniverse2024 Roman HLTDS images. The OpenUniverse2024 HLTDS covers 12 $\deg^2$ of simulated sky in a single deep tier, overlapping the LSST ELAIS-S1 Deep Drilling Field. It is inclusive of all seven Roman photometric bands and detector effects that are discussed in OpenUniverse2024. The PSF in the OpenUniverse2024 simulations does not change over time.

In this work, we use 10 pointings (full detector array) from each band for a total of 180 images (single sensor chip assembly, SCA) per band. Pointings are chosen such that each contains a particular coordinate location and therefore have overlap. Across the 10 pointings in each band, there are approximately $5.0 \times 10^4$ stars per band (including duplicates due to overlap) in the simulation catalogs.

[6] Code associated with this analysis can be found in the repository at https://github.com/Roman-Supernova-PIT/pub-aldoroty-2025. The PSFs themselves can be found on Zenodo: doi:10.5281/zenodo.15609513 (L. Aldoroty et al. 2025).

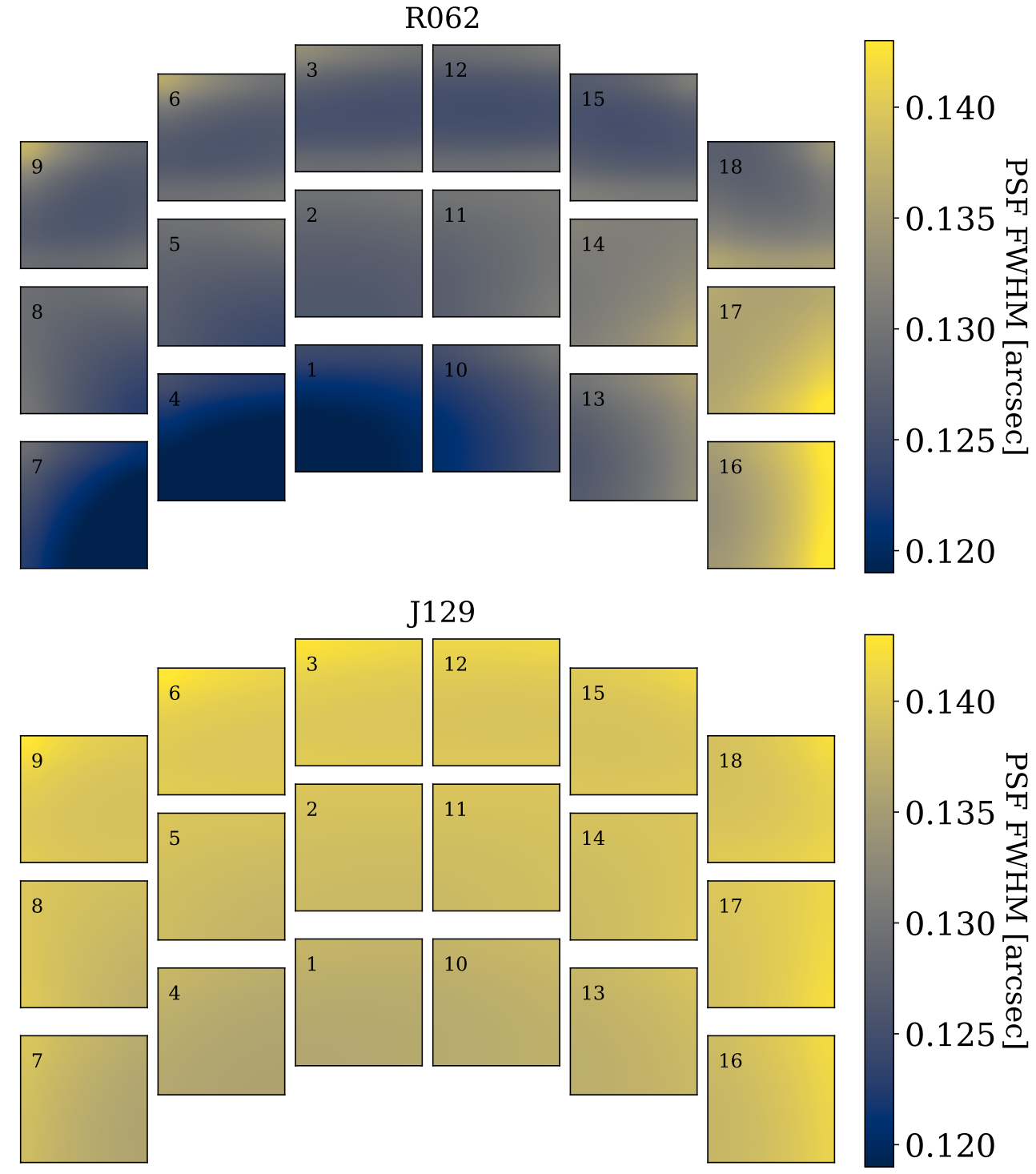

**Figure 1.** Focal plane variation of the PSF FWHM for different Roman bandpasses, assuming a flat spectral energy distribution. The shown positional dependence on the focal plane arises from optical aberrations alone and does not include any detector effects. The PSFs include the pixel response at the native scale but are oversampled by a factor of 8 in order to measure the sizes. The sizes are measured using adaptive weighted moments in GalSim. This plot is formatted to reflect Roman's detector shape and arrangement of all 18 SCAs.

**Table 1**
Grid Specifications Used in This Work

| $N_{grid}$ | $\Delta_{px}$ | $N_{stack}$ |
|---|---|---|
| 1 | 4088 | 3 |
| 2 | 2044 | 13 |
| 4 | 1022 | 50 |
| 8 | 511 | 250 |

### 2.2. PSF Generation

Roman's PSF is expected to vary significantly across the focal plane (Figure 1 and S. Casertano et al. 2021). The PSF used in the Roman OpenUniverse2024 simulations is based on ray tracing through an optical model of the telescope in order to build the most realistic possible PSFs. These PSFs have a varying FWHM across the focal plane (Figure 1) before detector effects are included. The FWHM of the PSFs that emerge from only optical effects varies by at least ∼1% and up to ∼10% for a single SCA (see Figure 1 for the arrangement of all 18 Roman SCAs) due to optical assembly alone. Thus, in order to make photometric measurements with the precision required for supernova cosmology with Roman, using the field stars from an entire image to generate a single PSF for use everywhere on the image introduces too much systematic error. In this work, we divide each SCA into a square grid with $N_{grid}$ elements on each side, where each grid element has a width in pixels $\Delta_{px}$ (Table 1). We generate a PSF for each grid element, which may be used on objects found within that

**Table 2**
Results for Photometric and Positional Precision

| $N_{grid}$ | $\Delta_{px}$ | R062 | Z087 | Y106 | J129 | H158 | F184 | K213 |
|---|---|---|---|---|---|---|---|---|
| | | | | $\hat{\sigma}_{(F_{fit}-F_{truth})/F_{truth}}$ | | | | |
| 1 | 4088 | 0.0158 | 0.0168 | 0.0120 | 0.0096 | 0.0091 | 0.0077 | 0.0092 |
| 2 | 2044 | 0.0139 | 0.0147 | 0.0103 | 0.0078 | 0.0086 | 0.0066 | 0.0082 |
| 4 | 1022 | 0.0132 | 0.0146 | 0.0100 | 0.0076 | 0.0085 | 0.0065 | 0.0081 |
| 8 | 511 | 0.0122 | 0.0140 | 0.0098 | 0.0074 | 0.0082 | 0.0062 | 0.0073 |
| | | | | $\hat{\sigma}_{x_{truth}-x_{fit}}$ | | | | |
| 1 | 4088 | 0.0933 | 0.0521 | 0.0277 | 0.0166 | 0.0124 | 0.0087 | 0.0119 |
| 2 | 2044 | 0.0948 | 0.0528 | 0.0266 | 0.0151 | 0.0117 | 0.0089 | 0.0098 |
| 4 | 1022 | 0.0956 | 0.0522 | 0.0269 | 0.0154 | 0.0115 | 0.0090 | 0.0100 |
| 8 | 511 | 0.0955 | 0.0527 | 0.0269 | 0.0149 | 0.0117 | 0.0086 | 0.0103 |
| | | | | $\hat{\sigma}_{y_{truth}-y_{fit}}$ | | | | |
| 1 | 4088 | 0.0546 | 0.0355 | 0.0265 | 0.0160 | 0.0108 | 0.0092 | 0.0109 |
| 2 | 2044 | 0.0540 | 0.0332 | 0.0238 | 0.0150 | 0.0108 | 0.0090 | 0.0109 |
| 4 | 1022 | 0.0545 | 0.0340 | 0.0240 | 0.0154 | 0.0109 | 0.0094 | 0.0107 |
| 8 | 511 | 0.0540 | 0.0339 | 0.0242 | 0.0154 | 0.0111 | 0.0104 | 0.0106 |

**Note.** Figure 2 corresponds to the values in the $N_{grid} = 8$ rows. Center: $\hat{\sigma}$ of the difference between truth catalog pixel coordinates and fit pixel coordinates, showing the precision of the recovered pixel coordinates in the $x$-axis by number of grid elements ($N_{grid}$) in each band. Bottom: the same as the center section but for the $y$-coordinates. The same figures correspond to this section of the table as the center section. All values in this table correspond to stars with $19 < m_{truth} < 21.5$.

region on an SCA. The total number of PSFs is $18 \times N_{grid}^2$, where 18 is the number of SCAs in the Roman detector.

J. Anderson & I. R. King (2000) recommend that each ePSF be based on approximately 100 stars. For a full $4088 \times 4088$ chip, the OpenUniverse2024 HLTDS simulations contain 150–200 stars that fall within the boundaries of the SCA (i.e., the star's pixel coordinates are between 0 and 4088 px in the simulated truth catalog for a given image). Of these, only 15–30 stars are neither saturated nor too dim (in this work, $19 < m_{truth} < 21.5$; see Section 3.1) to use for building an ePSF. This quantity only decreases for smaller grid elements. Therefore, a single grid element from an HLTDS image does not contain enough stars to generate an accurate and precise PSF, even if the full frame is used ($1 \times 1$). However, for the OpenUniverse2024 simulations, the PSF is solely a function of the simulated instrumentation itself and does not vary over time (Section 3.5.2). Thus, for each grid element, we use stars that fall within its defined region on the Roman detector collected from $N_{stack}$ total images (Table 1), chosen such that each PSF is based on approximately 100 stars (J. Anderson & I. R. King 2000). Stamp cutouts of these stars, which have been restricted to $19 \leqslant m_{truth} \leqslant 21.5$ in all bands, are used to construct an ePSF in `photutils` (J. Anderson & I. R. King 2000; J. Anderson 2016; L. Bradley et al. 2024). The final ePSFs are oversampled by a factor of 3. We use the simulated ("truth") catalog pixel coordinates as the initial conditions for star positions and aperture photometry from these known positions as the initial guess for flux. The output PSF is saved and applied to objects that fall within its designated region. We do not apply bilinear interpolation between PSFs to approximate the PSF at the specific location of the star (J. Anderson & I. R. King 2000) in order to more rigorously test the spatial variation of the Roman PSF. We do not use the same observations that were used to generate the PSFs in the final results of this work. However, we do use the same stars that were used to build each ePSF to calculate a zero point for each grid element, such that the median of $F_{truth} - F_{fit}$ is, by definition, zero.

### 2.3. Photometry

We apply the methodology described in this section to the chosen 10 pointings in each band (Section 2.1). The photometric measurements themselves are done by passing the ePSFs described in Section 2.2 to `photutils` (L. Bradley et al. 2024). We also provide the truth catalog coordinates of all stars in the image as initial conditions; however, we allow the coordinates to vary as fit parameters. We note that we are only able to set the true coordinates as initial conditions because we are analyzing simulated images. Although this is unrealistic, because the coordinates vary as fit parameters, the results are the same as if we had used coordinates from detection software like Source Extractor (E. Bertin & S. Arnouts 1996).

### 2.4. Source Detection

To test detection efficiency, we use Source Extractor (E. Bertin & S. Arnouts 1996) with default settings. Source Extractor determines an object to be "detected" if there is a group of adjacent pixels above a minimum size with values above a chosen detection threshold.

The output Source Extractor catalog is then cross-matched with the OpenUniverse2024 truth catalog by coordinate proximity. We require that detected stars must be within $0\overset{''}{.}1$ ($\sim$0.91 px) of the truth catalog coordinates in order to be matched.

## 3. Results and Discussion

### 3.1. Quantification of Stellar Measurement Precision and Accuracy

In order to characterize photometry, we follow a series of standard procedures to evaluate precision and accuracy. First,

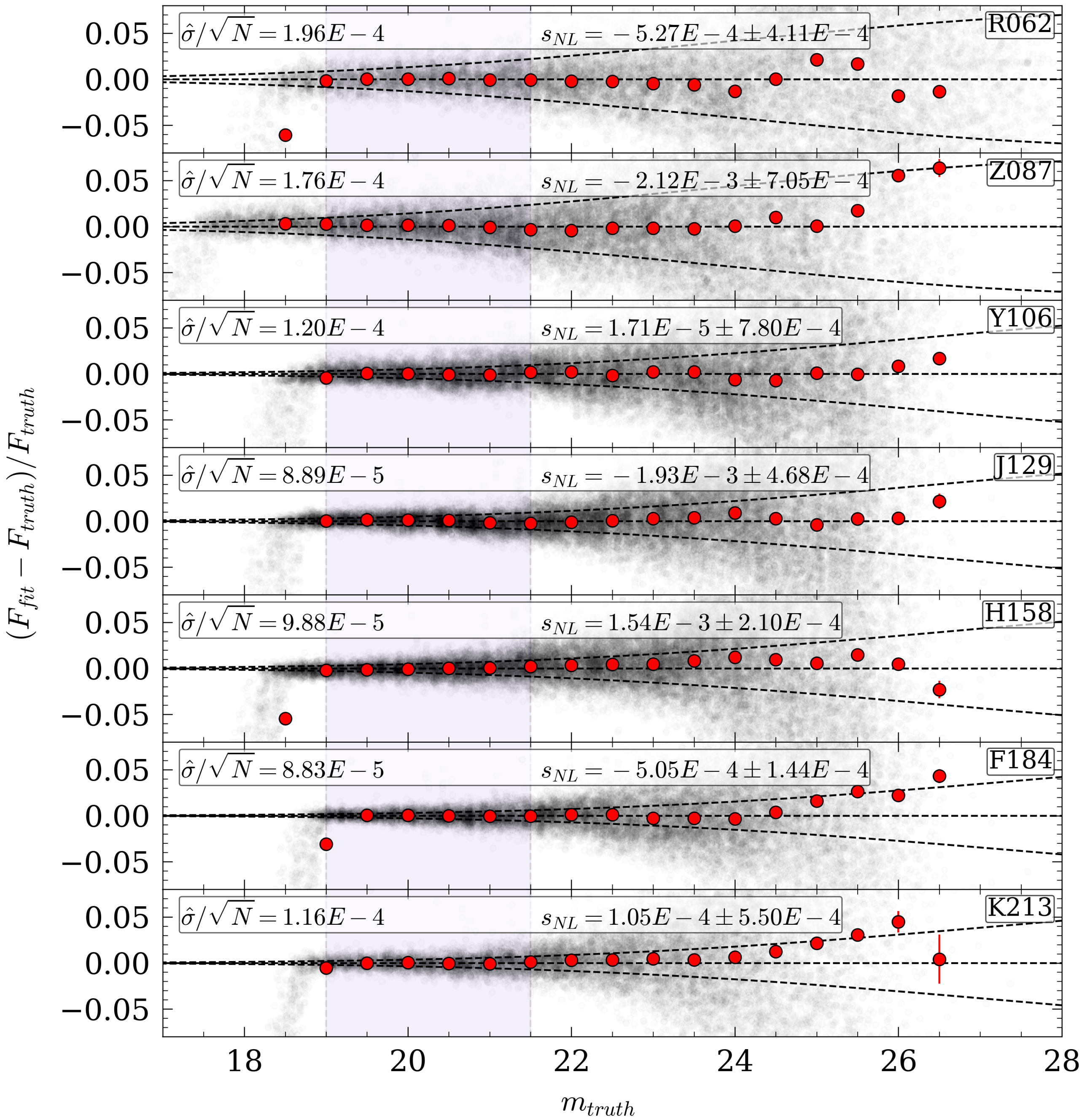

**Figure 2.** Photometric recovery, in terms of fractional flux difference, for field stars in one pointing from OpenUniverse2024. Black points represent each individual star. Red points are $(F_{\rm truth} - F_{\rm fit})/F_{\rm truth}$ binned every 0.5 mag, with error bars showing the standard deviation of that bin. Field stars are located based on truth coordinates, and the PSF is constructed using `photutils.psf.EPSFBuilder` (J. Anderson & I. R. King 2000; J. Anderson 2016; L. Bradley et al. 2024) for an $8 \times 8$ grid. The shaded purple region, $19 \leqslant m_{\rm truth} \leqslant 21.5$, shows the magnitude range over which stars were selected for PSF construction, as well as zero-point calculation. Curved black dashed lines represent Poisson noise for the stars only (i.e., background is excluded). $\hat{\sigma}$ is the modified median absolute deviation assuming a normal distribution (Equation (1)), and $s_{\rm NL}$ is the slope of the binned data (nonlinearity) in the shaded purple region excluding the bin at $m_{\rm truth} = 19$. The median of all panels is zero by design; the flux measurements are zero pointed to the truth catalog for $19 \leqslant m_{\rm truth} \leqslant 21.5$.

we develop a library of ePSFs for the OpenUniverse2024 simulations. The library contains ePSFs generated for grids of different sizes ($1 \times 1$, $2 \times 2$, $4 \times 4$, and $8 \times 8$) that can be directly applied to objects that fall within a particular ePSF's prescribed area on the detector. We base all our statistics and our ePSFs on stars with $19 < m_{\rm truth} < 21.5$ because we find that the magnitude range $19 < m_{\rm truth} < 21.5$ corresponds to stars that are neither saturated nor sky noise limited in all bands. Brighter than $m_{\rm truth} \approx 19$, we find a "tail" that reflects the saturated stars having $m_{\rm fit} > m_{\rm truth}$. Dimmer than $m_{\rm truth} \approx 21.5$, photometric measurements begin to become noisy.

After the ePSFs are generated, we test their suitability for recovering input fluxes by fitting them to stars that were not used to generate the ePSFs. We summarize the results in Table 2 and Figures 2 and 3. We calculate a modified median absolute deviation ($\hat{\sigma}$) of the fractional flux difference between truth and fit fluxes (Section 2.3) for these stars,

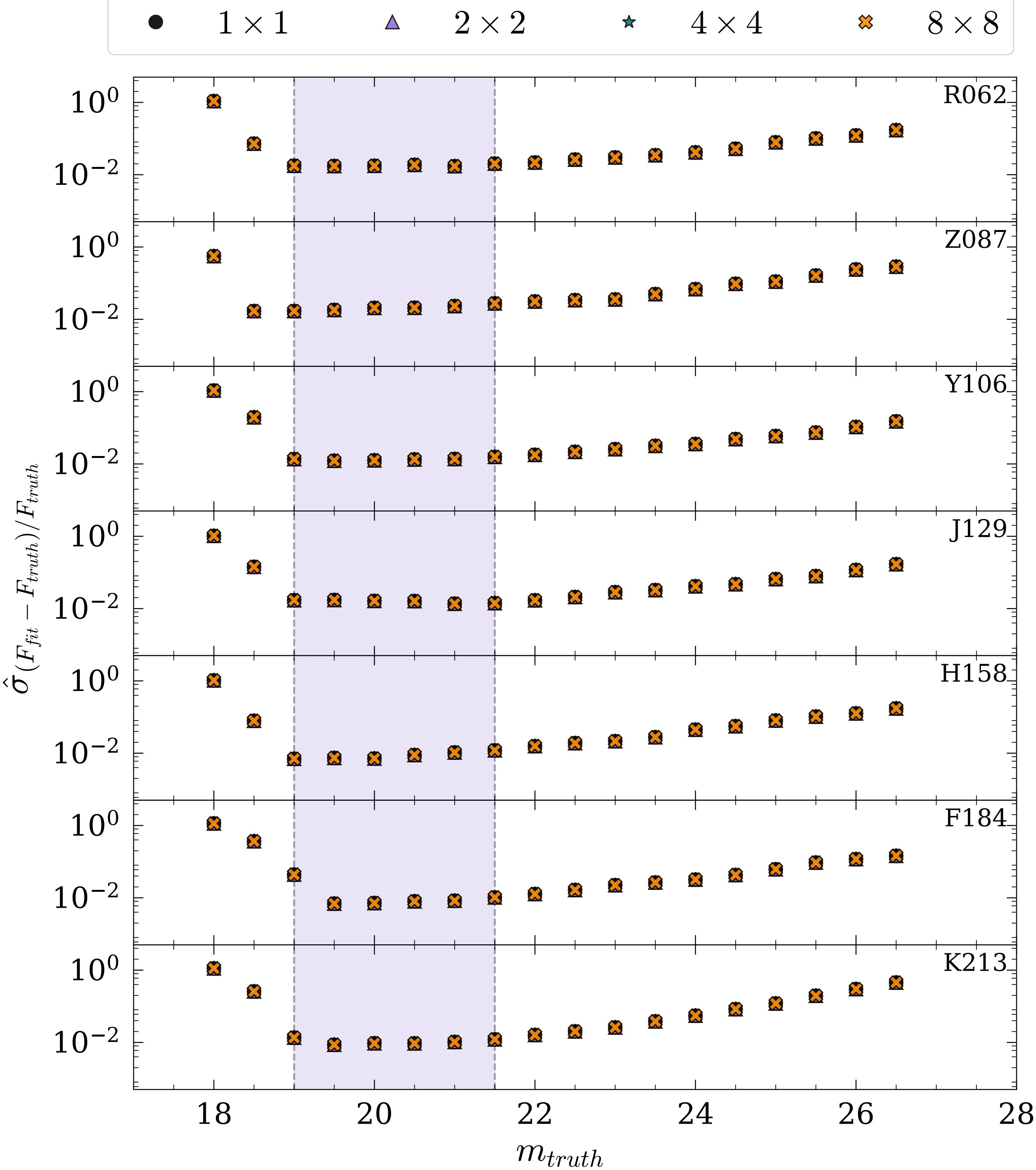

**Figure 3.** Binned modified median absolute deviation ($\hat{\sigma}$) assuming a Gaussian distribution of the fractional flux residuals for all grid sizes. Each color and marker style represents a different grid size (see legend). The shaded purple region is $19 < m_{\text{truth}} < 21.5$.

$(F_{\text{truth}} - F_{\text{fit}})/F_{\text{truth}}$, defined as

$$\hat{\sigma} = 1.48 \times \text{median}\left(\left|\frac{F_{\text{truth}} - F_{\text{fit}}}{F_{\text{truth}}}\right|\right), \tag{1}$$

where the factor of 1.48 comes from an assumption that $(F_{\text{truth}} - F_{\text{fit}})/F_{\text{truth}}$ has a Gaussian distribution. We show binned $\hat{\sigma}$ versus $m_{\text{truth}}$ in Figure 3.

### *3.2. Spatial Dependence, Color Dependence, and Magnitude Dependence (Nonlinearity)*

Using the $8 \times 8$ grid ($N_{\text{grid}} = 8$), we show that our ePSF library is able to achieve the expected <1% flux precision in all bands except R062 and Z087. For the $8 \times 8$ grid, $\hat{\sigma}_{\text{R062}} = 0.0122$ and $\hat{\sigma}_{\text{Z087}} = 0.0140$. We also demonstrate the importance of considering the spatial variation of the

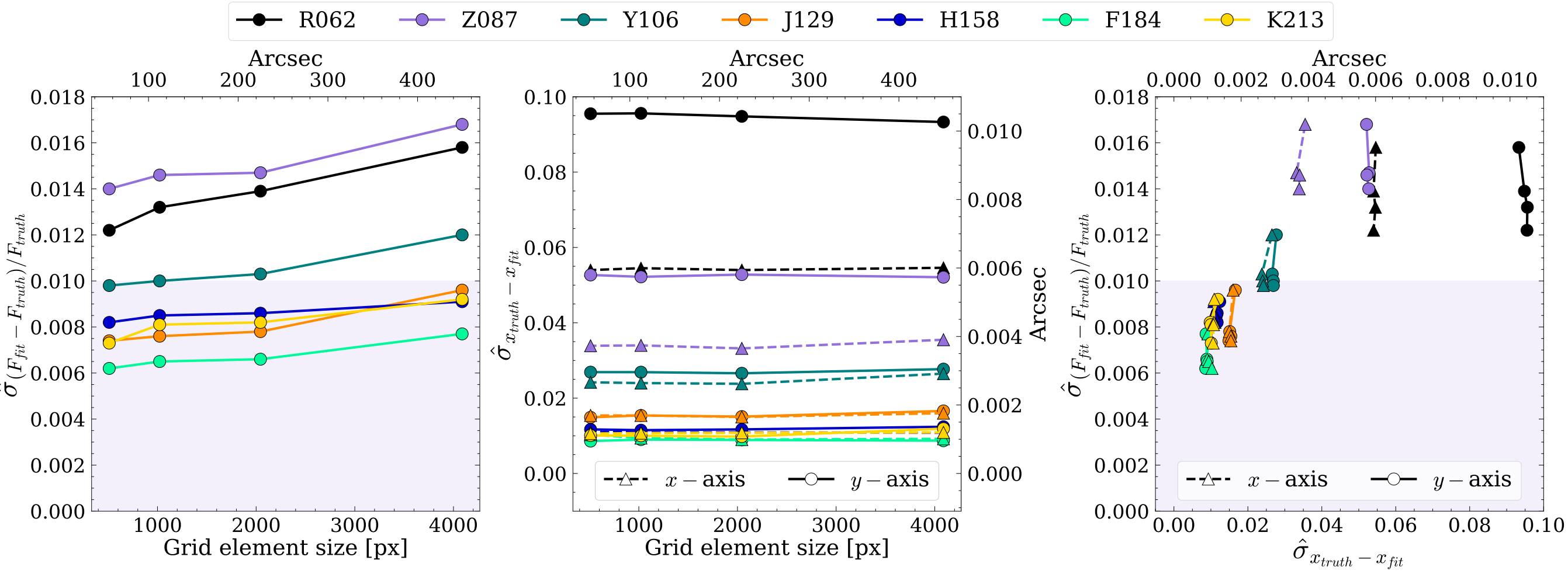

**Figure 4.** Left: the modified median absolute deviation ($\hat{\sigma}$), assuming a normal distribution, of the fractional difference between truth catalog flux and the fit PSF photometry flux as a function of grid element size in pixels. Center: $\hat{\sigma}$, assuming a normal distribution, of the difference between truth catalog pixel coordinates and fit pixel coordinates. Right: $\hat{\sigma}$ of the fractional flux differences from the left panel compared to the pixel locations in the center panel. Circular markers with solid lines correspond to the *x*-coordinates, and triangles with dashed lines correspond to *y*-coordinates. The shaded purple region represents <1% flux precision. This figure corresponds with the results in Table 2.

Roman PSF; as the grid becomes finer, flux precision improves. In particular, if we compare the results from using stars from full images ($1 \times 1$ grid) to the $8 \times 8$ grid, fractional flux precision improves by a range of 9%–30%, without obvious dependence on filter (Table 2; Figure 4).

All bands show a flat "noise floor," such that even for brighter (unsaturated) objects, $\hat{\sigma}$ does not continue to decrease (Figure 3). Additionally, for the $8 \times 8$ grid, all bands except R062 and Z087 have $\hat{\sigma}$ greater than the Poisson noise, $\sigma_\lambda$, at all magnitudes (Figure 5). In particular, we note that in the $19 < m_{\rm truth} < 21.5$ region, $\hat{\sigma}$ represents a larger portion of systematic uncertainty when compared to $\sigma_\lambda$, especially for brighter $m_{\rm truth}$. This is notable because $\sigma_\lambda$ increases for brighter objects. More work is required to identify the source of these systematics.

While analysis of Roman measurements will have to contend with both classical nonlinearity and count-rate nonlinearity of the detectors, these effects are not simulated in OpenUniverse2024. However, nonlinearity may be introduced by errors in photometric recovery. For the $8 \times 8$ grid of ePSFs, we measure the nonlinearity slope ($s_{\rm NL}$) of the fractional flux recovery in the nonsaturated, sky-noise subdominant region ($19 < m_{\rm truth} < 21.5$; Figure 2). Given our calculated values of $s_{\rm NL}$, we find that $\hat{\sigma}/\sqrt{N}$, where $N$ is the number of stars in the sky-noise subdominant region. We measure $|s_{\rm NL}| < 1.93 \times 10^{-3}\ {\rm dex}^{-1}$ in all bands (Figure 2). This value is larger than the requirement for the detector count-rate nonlinearity (S. Casertano et al. 2021) of <0.3% over 11 mag due to detector effects. Given that this requirement stemmed from cosmology forecasts, further studies should attempt to reduce any possible nonlinearity due to photometry.

We analyze the impact of color dependence on our results in Figure 6. We bin our data every 0.05 mag by $\rm J129_{fit} - H158_{fit}$ color and measure the slope, $s_{\rm color}$, of the binned data from $-0.15 < \rm J129_{fit} - H158_{fit} < 0.15$. We find a significant slope ($\sim 8\sigma$) in R062 but smaller slopes in the other bands both in size and significance ($<3.5\sigma$). Similarly, we can compare $s_{\rm color}$ with the photometric floor of $\hat{\sigma}_{(F_{\rm fit} - F_{\rm truth})/F_{\rm truth}}$ found in Figures 2 and 3 and again find R062 to be at the high end for relative slopes. A possible implication is that a color-dependent PSF model is needed (M. Jarvis et al. 2021a), at least in R062, but potentially for other bands and should be studied in follow-up work.

### 3.3. Limitations

#### 3.3.1. Pixel-phase Bias

It can be difficult to ascertain the position of point sources in undersampled images, often leading to a bias in measured position and therefore also inaccurate photometric measurements (T. R. Lauer 1999). If the PSF is known precisely, then determining the position of stars is possible because it is known how much flux falls in the surrounding pixels if the star is perfectly centered on a given pixel. J. Anderson & I. R. King (2000) show that this positional error, also called the "pixel-phase error," can be the result of an imprecise PSF model in the HST WFC images.

Dithered observations of the same stars are required to address the "pixel-phase bias" described in J. Anderson & I. R. King (2000). The "pixel-phase bias" is a result of degeneracy between the star's position and the shape of the PSF itself. We note that the `photutils` (L. Bradley et al. 2024) implementation of the ePSF method (J. Anderson & I. R. King 2000; J. Anderson 2016) does not explicitly handle dithered observations and, therefore, repeated stars. Although we have attempted to alleviate this issue with degeneracy by using the true coordinates of the stars as initial conditions, there is no convenient way in `photutils` to set these positions as fixed, which would eliminate the degeneracy for this particular application (i.e., simulated stars with known coordinates) and therefore eliminate the need for dithered observations. For future work involving generation of new ePSFs, we would seek an implementation of this technique that includes the capability to specify known coordinates, such as `PIFF` (M. Jarvis et al. 2021a, 2021b; T. Schutt et al. 2025).

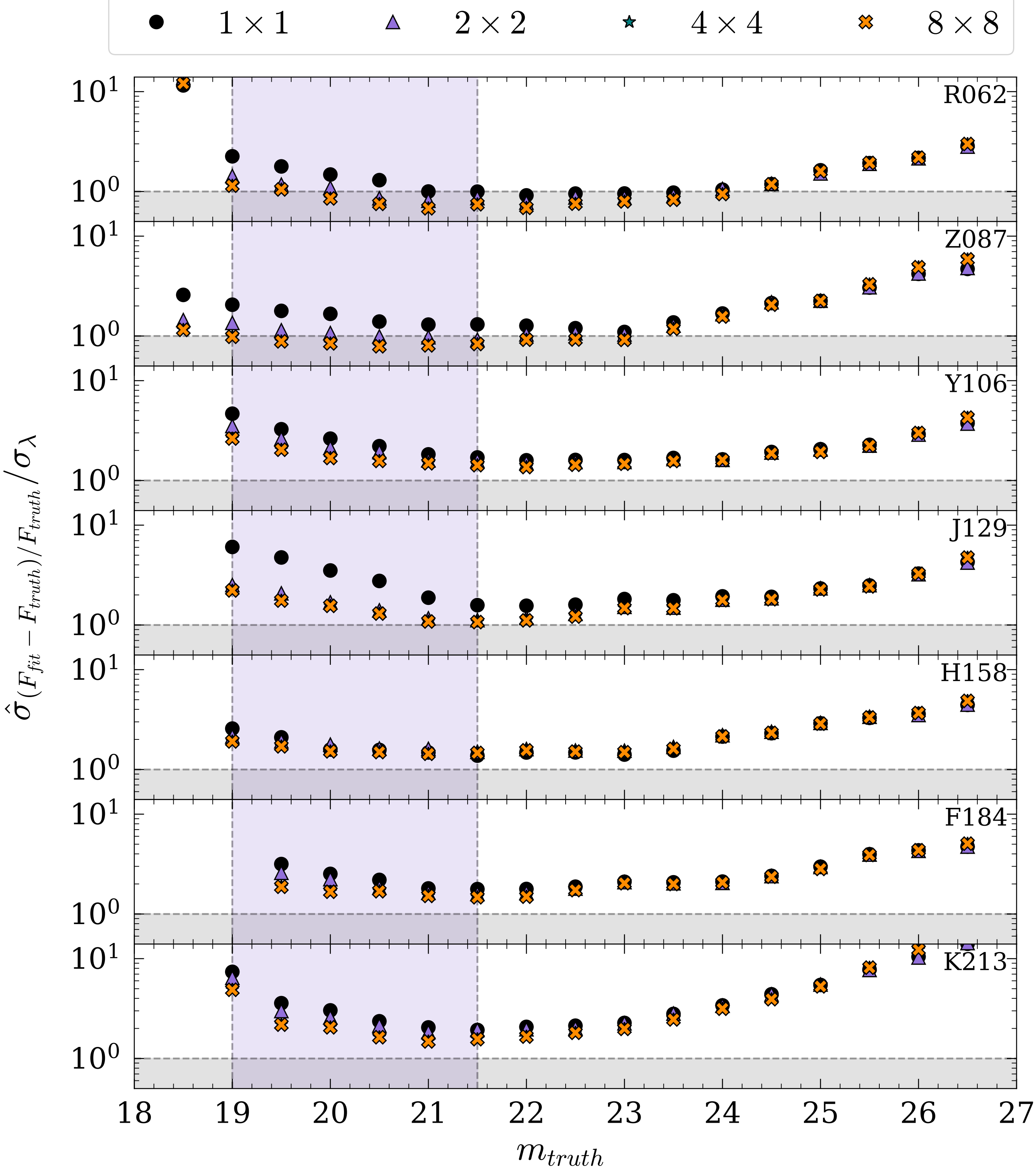

**Figure 5.** The vertical axis values in Figure 3 divided by Poisson noise ($\sigma_\lambda$). Each color and marker style represents a different grid size (see legend). The shaded purple region is $19 < m_{\rm truth} < 21.5$, and the shaded gray region is where $\hat{\sigma}/\sigma_\lambda \geqslant 1$.

We note that despite the inability to fix the locations of the stars for generating the ePSF, we were able to recover the locations of the measured stars to within less than 0.03 px in all cases except for the R062 and Z087 filters (Table 2, center; Figure 4, right)—around $\frac{1}{3}\times$ Roman's pixel scale ($0\overset{''}{.}11$ px$^{-1}$). As expected, recovered magnitude precision is loosely correlated with recovered coordinate precision (Figure 4, right). Coordinate recovery precision does not appear to improve with the number of grid elements like $\hat{\sigma}_{(F_{\rm fit}-F_{\rm truth})/F_{\rm truth}}$ does (Figure 4, center). In the R062 and Z087 filters, the fit pixel location is especially biased with respect to the true catalog coordinates. This is unsurprising because the images in these filters are more severely undersampled than the other five, which means it is more difficult to determine the centroid of a particular object. Additionally, these PSFs are the two most asymmetric of the seven in this work (Section 3.3.3). We also see that $\hat{\sigma}_{y_{\mathrm{truth}} - y_{\mathrm{fit}}}$ is consistently lower than the $\hat{\sigma}_{x_{\mathrm{truth}} - x_{\mathrm{fit}}}$ (Figure 4, center), which we attribute to PSF ellipticity and discuss in more depth in Section 3.3.3.

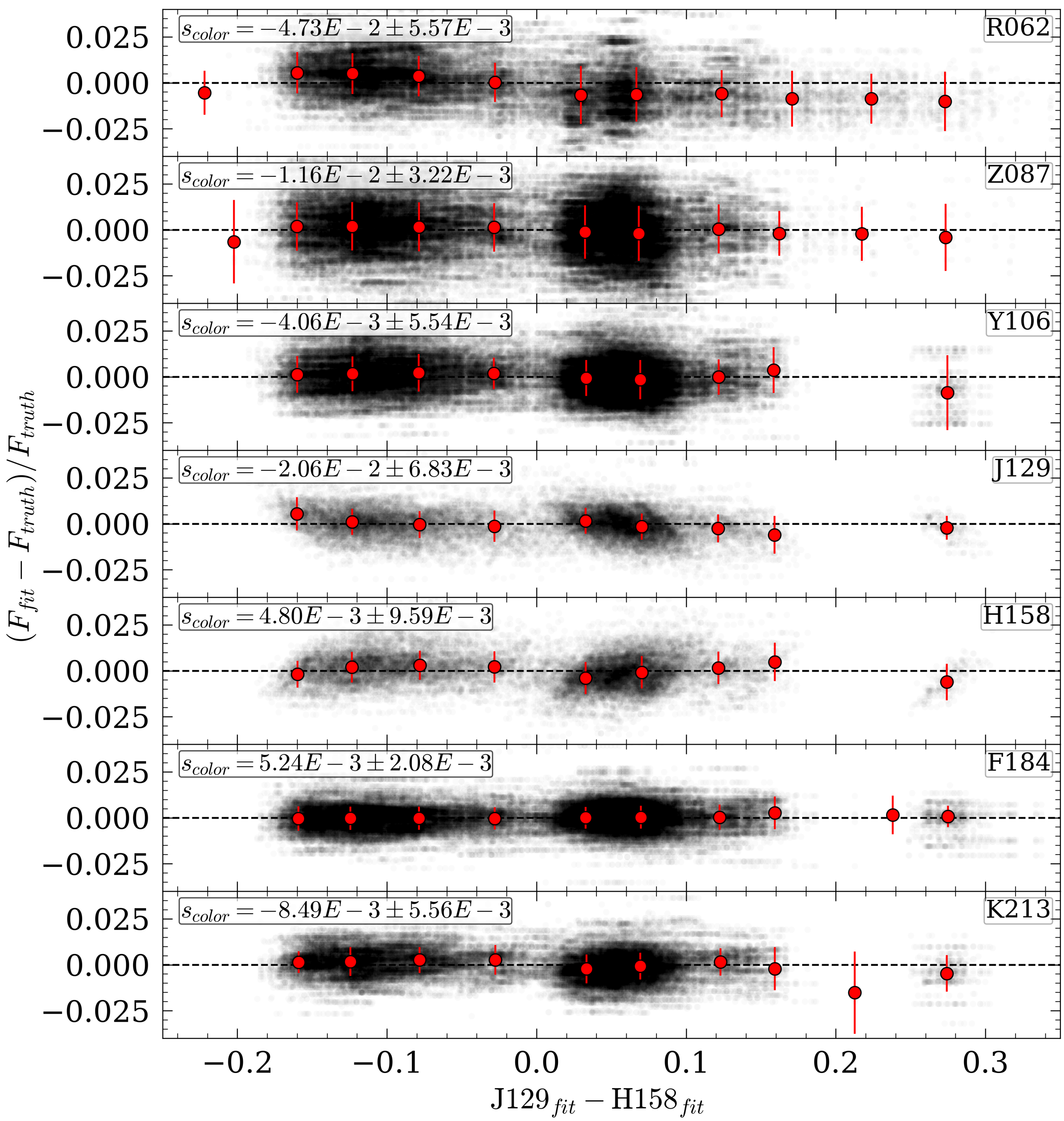

**Figure 6.** Chromatic dependence of $(F_{\mathrm{fit}} - F_{\mathrm{truth}})/F_{\mathrm{truth}}$ for stars with $19 < m_{\mathrm{truth}} < 21.5$. Small black points are individual measurements. Larger red points are the data binned every 0.05 mag, with the position of the red point at the mean of of $\mathrm{J129}_{\mathrm{fit}} - \mathrm{H158}_{\mathrm{fit}}$ in that bin. The fit slope of the red points between $-0.15 < \mathrm{J129}_{\mathrm{fit}} - \mathrm{H158}_{\mathrm{fit}} < 0.15$ with its associated error is reported as $s_{\mathrm{color}}$ in the upper-left corner for each panel.

### 3.3.2. PSF Asymmetry

We find that in some cases, the fit position of stars from PSF fitting (Section 2.3) is systematically offset from the truth coordinates in accordance with observing filter and which SCA it falls on (Table 2; Figure 7). We attribute this to PSF asymmetry. The PSFs from the OpenUniverse2024 simulations, which are produced based on the optical systems of the real Roman Space Telescope, reflect the asymmetry expected in the PSFs; in the left column of Figure 8, we show the OpenUniverse2024 GalSim PSFs prior to interaction with the detector. These PSFs show significant structure and asymmetry. The center column of Figure 8 shows the same PSFs after convolution with Roman's detector pixels, which obscures the structure shown in the left column. The ePSFs generated in this work are shown in the right column for comparison to the "true" PSFs. We quantify this asymmetry by dividing each PSF image into a 2 × 2 grid and defining an asymmetry index, $\alpha$, which is a triangular operation over this grid written as

$$\alpha = \frac{1}{4} \sum_{\substack{i,j=[0,\,3] \\ i<j}} \frac{|A_i - A_j|}{A_i + A_j}, \tag{2}$$

where $A$ is the sum over the pixels in each quadrant. We compute $\alpha$ at the center of each SCA in all bands (Appendix A, Table 4). The left panel of Figure 9 shows that fractional flux scatter is highly correlated with PSF asymmetry, regardless of $N_{\rm grid}$. We discuss this result further in Section 3.3.3.

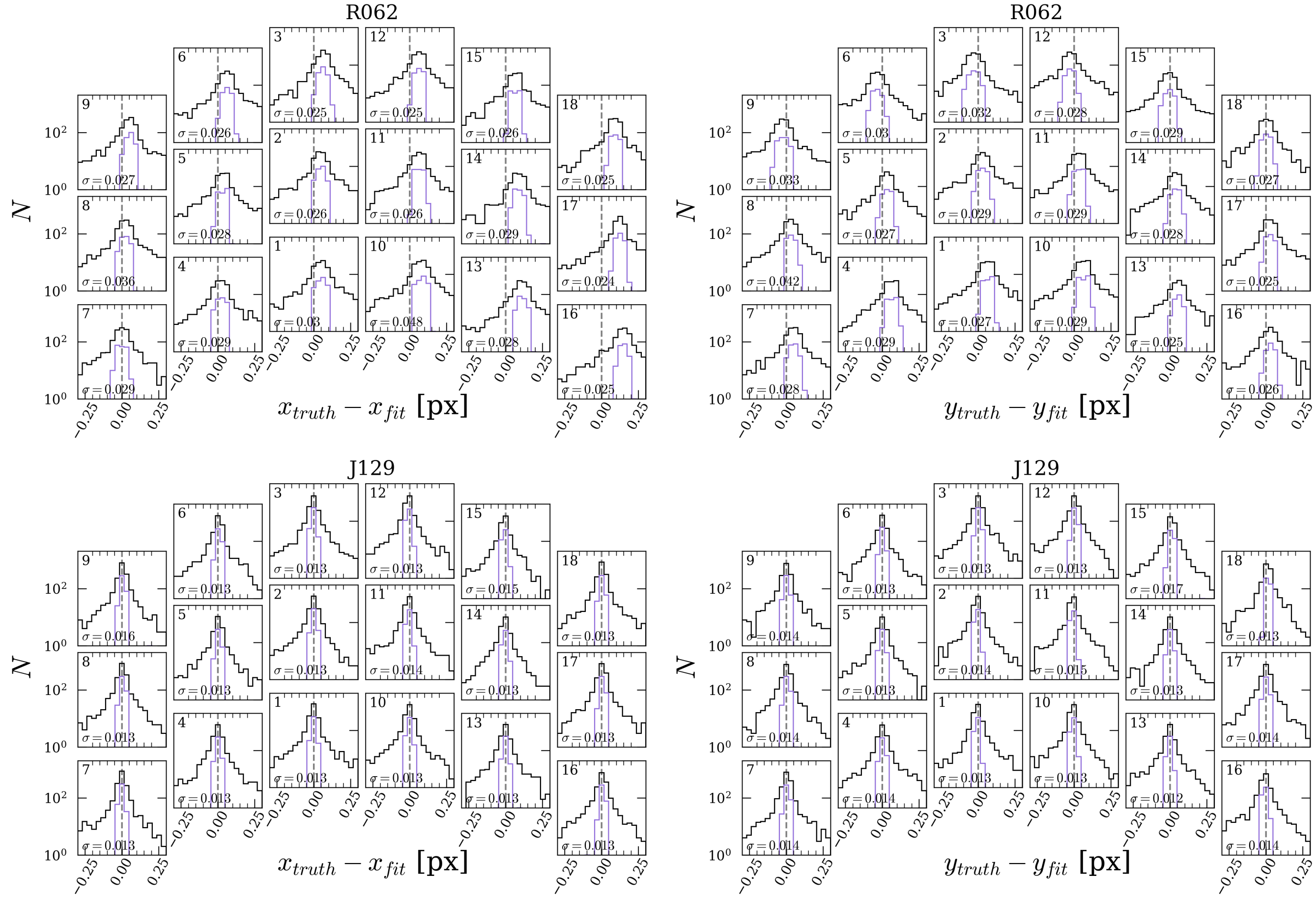

**Figure 7.** Top: star location offsets in pixel coordinates for an 8 × 8 grid, arranged by individual SCA. The black line is all stars, and the purple line is stars with $19 < m_{\rm truth} < 21.5$. Bottom: the same for the J129 band. This plot is formatted to reflect Roman's detector shape and SCA arrangement. $\sigma$ is the standard deviation of the purple line. We choose to show the R062 and J129 bands to illustrate this discussion point (Figures 4 and 10) because they have the largest and smallest differences, respectively, in their $x$- and $y$-coordinate recovery precision (Table 2).

For more asymmetric PSFs (R062, Z087, and Y106), we observe a systematic pixel offset in stars' fit coordinates compared to their catalog coordinates (Figure 8, left; Figure 7, top). This offset is not present in cases where the PSF is more symmetric; for the J129 and redder bands, the PSF does not show significant asymmetry (Figure 8), and we do not see a systematic offset in recovered coordinates when compared to the simulated coordinates (Figure 7).

Although our ePSFs show some asymmetry, it is not as pronounced as in the GalSim PSF (Figure 8). Thus, it is unsurprising that the fit coordinates are systematically offset from the simulated coordinates in areas where the simulations' input PSF is more asymmetric. This effect may also contribute to the increased scatter in fractional flux for the R062, Z087, and Y106 bands.

### 3.3.3. PSF Ellipticity

The Roman PSF is expected to have nonzero ellipticity. Like the FWHM, the effects of the ellipticity are expected to be more severe in bluer bands than redder bands (Figures 10 and 11) as it is a property associated with asymmetry. Ellipticity is quantified by the constants $e_1$ and $e_2$ (Appendix B). By definition, if $e_1 > 0$, the ellipse is elongated along the $x$-axis. Oppositely, if $e_1 < 0$, the ellipse is compressed along the $x$-axis. If $e_2 > 0$, the ellipse is stretched along $y = x$. Similarly, if $e_2 < 0$, it is stretched along $y = -x$. If $e_1 = e_2 = 0$, there is no stretch (i.e., the ellipse is a circle). If a PSF has nonzero ellipticity, it is expected that the distribution of recovered coordinates will be wider than for a PSF with zero ellipticity.

In this work, we find that the width of the distributions of recovered coordinate offsets is larger for SCAs with an offset (Figure 7). Our results reflect the expected relationship between ellipticity effects and distribution width; the widths of the recovered coordinate distributions are consistently larger for bluer bands than other bands (Figure 7). In particular, we draw attention to Figure 12; for PSFs that are more asymmetric (larger $\alpha$) and elliptical (larger $\sqrt{e_1^2 + e_2^2}$), the offsets in the corresponding SCA in Figure 7 are larger.

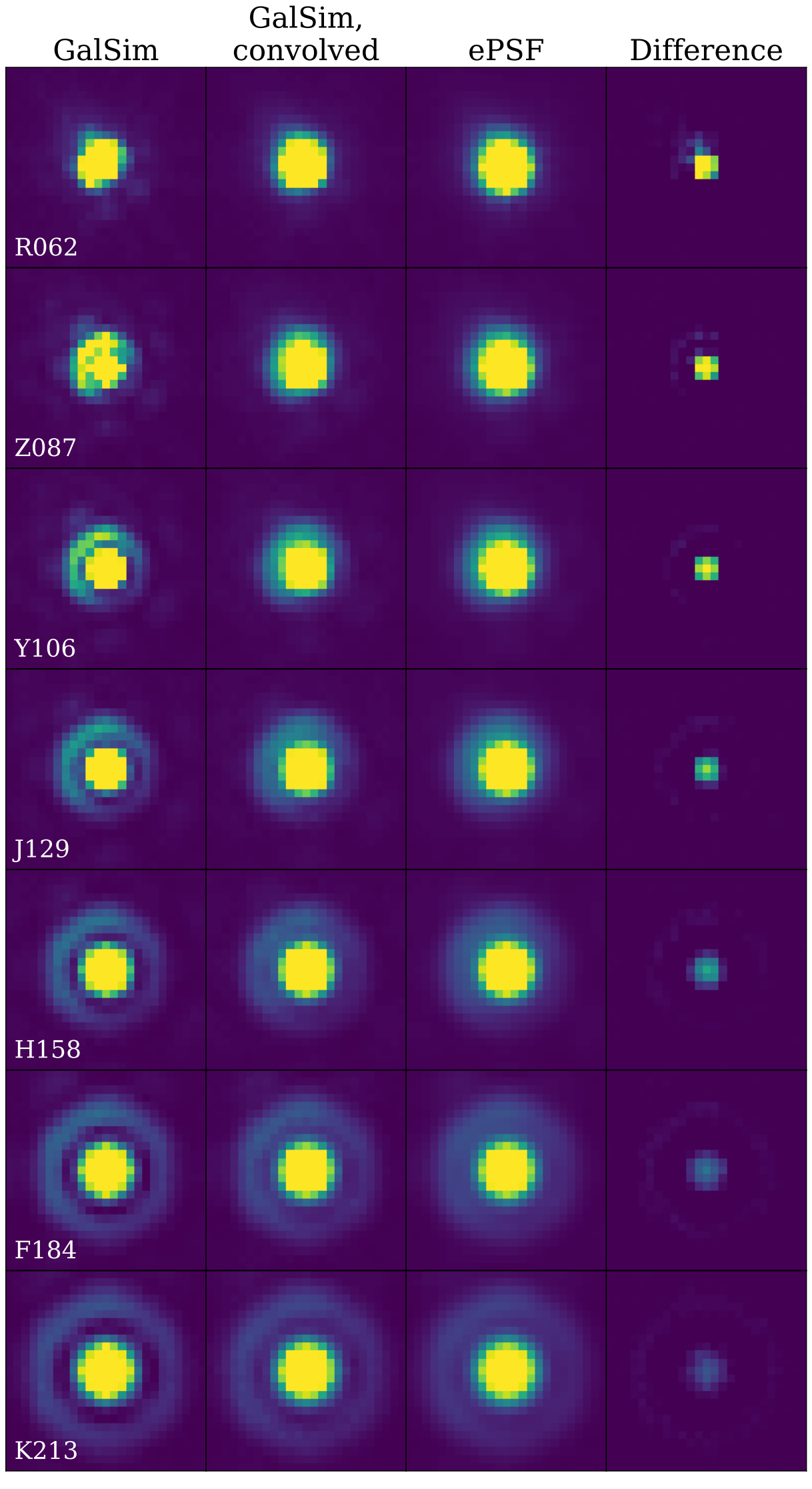

**Figure 8.** First (left) column: injected GalSim PSFs for the all Roman filters at the same location on the same SCA. Second column: GalSim PSFs from the first column, after convolution with the Roman pixel response. Third column: ePSFs from the 8 × 8 grid for the same location as the GalSim PSFs. Fourth column: the difference between the second and third columns. All PSFs are shown with the same color scale, and all have an oversampling factor of 3. The left three columns are normalized such that the values sum to 1. The right column is not normalized after subtracting the normalized third column from the normalized second column. Asymmetry in the displayed PSFs is caused by Roman's optical system.

Although ellipticity is a measurable portion of asymmetry, it does not account for all asymmetry. We note that the ellipticity of redder bands is negligible (Figure 9, right) despite the average asymmetry index $\langle\alpha\rangle$ still differing between the same bands. In other words, ellipticity does not significantly contribute to asymmetry for the H158, F184, and K213 bands, and their asymmetry is due to other properties.

### *3.4. Detection Efficiency*

We fit our binned detection efficiency curves (Figure 13) with a logistic function, defined in S. A. Rodney et al. (2014) as

$$\eta_{\text{det}}(m) = \eta_0 \times \left(1 + \exp\left(\frac{m - m_{50}}{\tau}\right)\right)^{-1}, \tag{3}$$

where $\eta_0$ is the maximum efficiency, $m_{50}$ is the magnitude where 50% of stars are recovered, and $\tau$ describes the exponential curve.

We find that we detect nearly all stars down to approximately 26th magnitude before a steep drop-off in all bands (Figure 13; Table 3). Our chosen range of neither saturated nor noisy stars ($19 < m_{\text{truth}} < 21.5$) fits well within the range where all stars are detected.

Although there are approximately $5 \times 10^4$ stars in the truth catalogs in each band, summed over all 10 pointings, a large portion of these drop away when objects that fall within SCA gaps, or are otherwise outside the detector, are excluded. There are around $3 \times 10^4$ remaining stars across all 10 pointings in each band after this exclusion, matching the quantities in the bottom row of Figure 13.

### *3.5. Considerations for Future Work*

#### *3.5.1. Dithering*

A major difference between the HLTDS and the dither patterns discussed in J. Anderson & I. R. King (2000) is that the HLTDS does not have explicit dithering because the planned rotations will introduce dither, whereas J. Anderson & I. R. King (2000) use an explicit $x$, $y$ dither pattern. In the typical translational dither pattern, the telescope is moved only slightly so that a given star's pixel coordinates on the detector are relatively near each other. For small rotations, we anticipate no reason that a rotational, rather than translational, dither pattern would pose a problem for addressing pixel-phase bias. However, with the planned rotational dither for the HLTDS, a single star will appear in many locations across the detector. Thus, its pixel coordinates in different exposures may be far apart. At each of these locations, the Roman PSF will be quite different. As we previously demonstrated, the varying ellipticity of the PSF means that it is harder to determine accurate and precise positions of objects at certain locations on the detector; however, not many repeated observations of the same star are needed to break the degeneracy between PSF shape and star location—J. Anderson & I. R. King (2000) states that only a 2 × 2 dither pattern is likely sufficient. Thus, results such as the ones described in this work will be important for determining how far apart these pixel coordinates can be in order to qualify as appropriate data for generating a PSF with minimal pixel-phase bias, in addition to deciding which specific SCAs (Section 3.3.3) are best for the task.

#### *3.5.2. Simulations versus Real Images*

This analysis is carried out on simulated images (OpenUniverse2024). Thus, not all real detector and instrumentation effects are accounted for in this work. For example, we assume that the instrumental PSF does not change over time. While this is an appropriate assumption to make for simulations, this is not the case for a real scenario. J. Anderson & I. R. King (2000) find that for HST, the PSF negligibly changes over the timescale over a few orbits. However, over a period of several years, this natural PSF evolution affects the pixel-phase error significantly. It is not currently known how much the

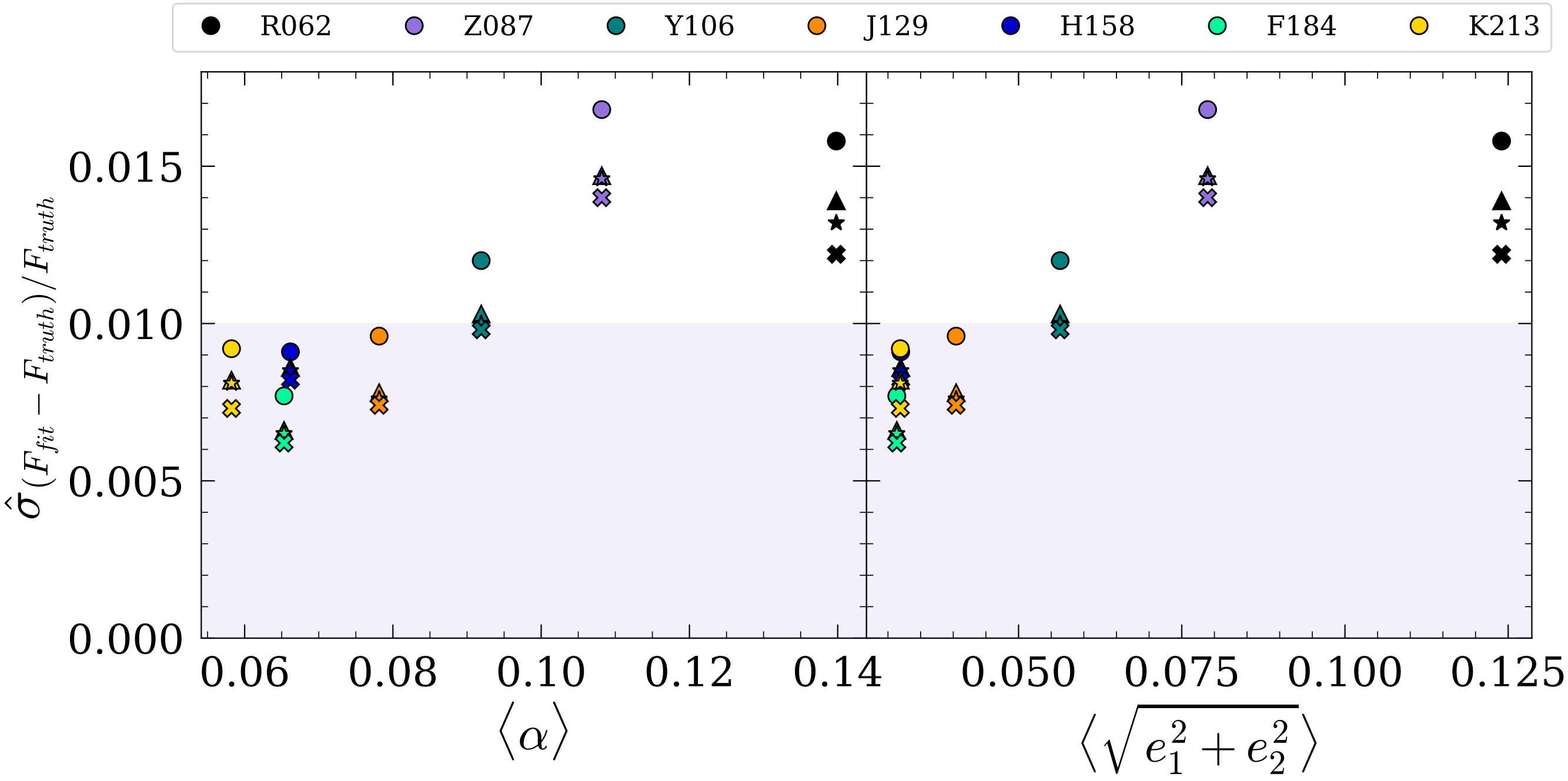

**Figure 9.** Modified median absolute deviation, $\hat{\sigma}_{(F_{\rm fit}-F_{\rm truth})/F_{\rm truth}}$, as a function of average asymmetry ($\alpha$, Equation (2)) per band (left) and average ellipticity (Appendix B) per band (right). Varying symbol shapes represent different grid sizes and are the same as in Figure 3. The shaded purple region represents <1% flux precision.

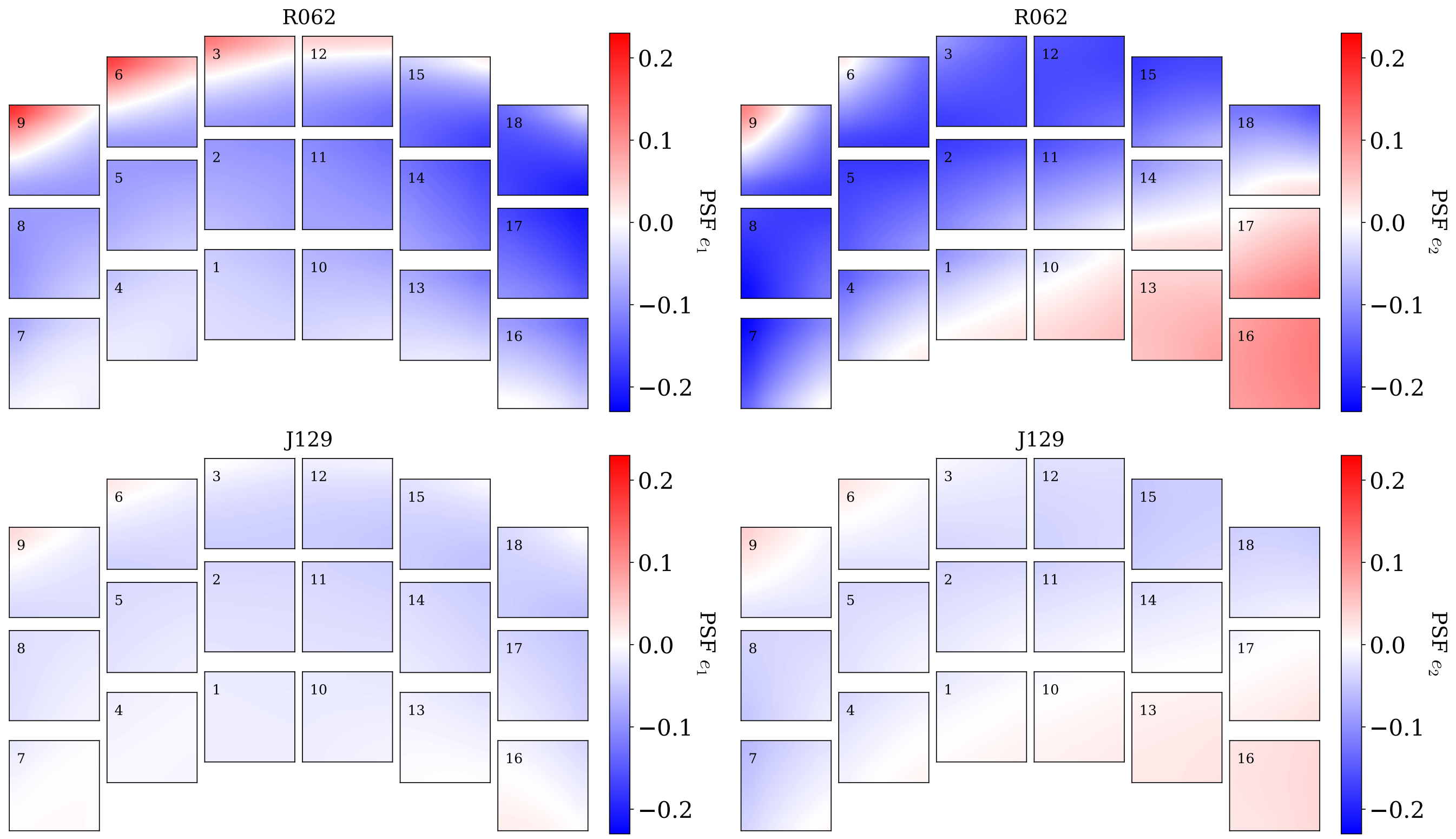

**Figure 10.** Top: ellipticity constants $e_1$ (left) and $e_2$ (right) of the injected GalSim PSF in the R062 band. Bottom: the same for the J129 band. This plot is formatted to reflect Roman's detector shape and SCA arrangement.

Roman PSF will change over time. Thus, it is unknown how often the official Roman ePSFs will need to be recalibrated. The current Roman WFI on-orbit calibration plan includes monitoring photometric "touchstone" fields, which will be observed regularly in order to monitor stability over time (S. Casertano et al. 2021).

### 3.5.3. Official Roman PSF Libraries

One of the differences between the ePSF library from this work and the ePSFs that the Roman Science Operations Center (SOC) will provide (S. Casertano et al. 2021) is that we apply our ePSFs directly to objects, whereas the ePSF library provided by the SOC will use interpolation between gridded ePSFs to retrieve the PSF for a particular location (as in J. Anderson & I. R. King 2000 and J. Anderson 2016). At this time, it is unknown if applying bilinear interpolation between our gridded ePSFs improves precision. We note that our $8 \times 8$ grid is the only grid size with an odd-valued $\Delta_{\rm px}$ (Table 1); given this as well as its superior performance compared to the other grid sizes, it is the best set of ePSFs to use on the OpenUniverse2024 simulations going forward.

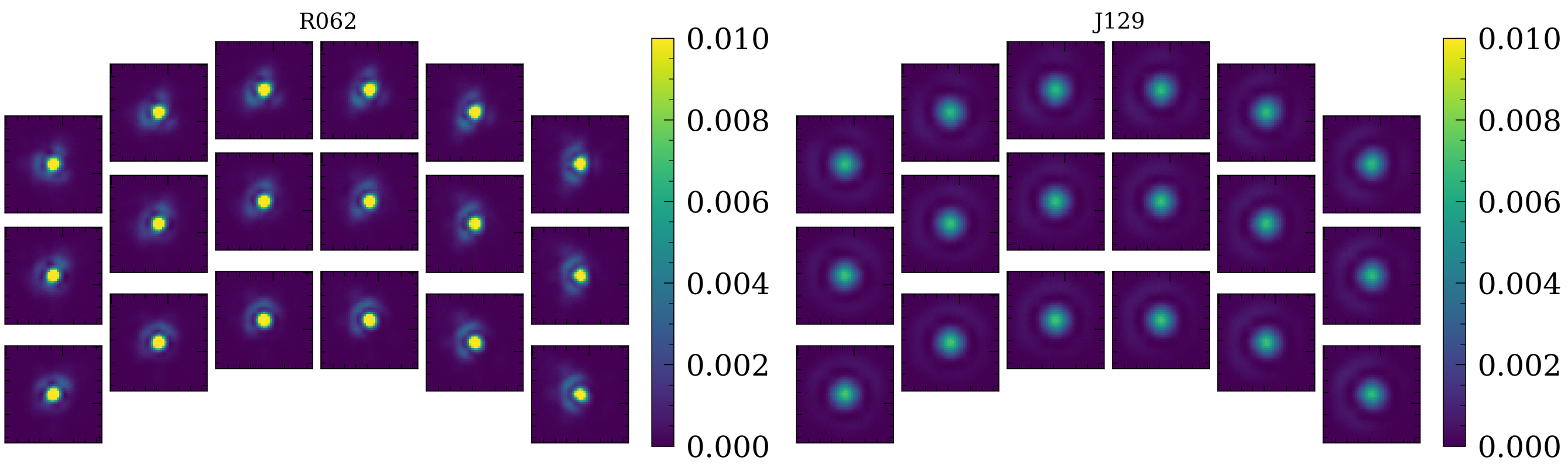

**Figure 11.** GalSim PSFs, unconvolved with the Roman pixel response, at the center of each SCA. All PSFs are oversampled by a factor of 9. This plot is formatted to reflect Roman's detector shape and SCA arrangement. Note that for the bottom two rows of the detector, the $x$-axis is defined such that the origin is at the bottom-right corner of the detectors, and positive $x$ points left (M. A. Troxel et al. 2023).

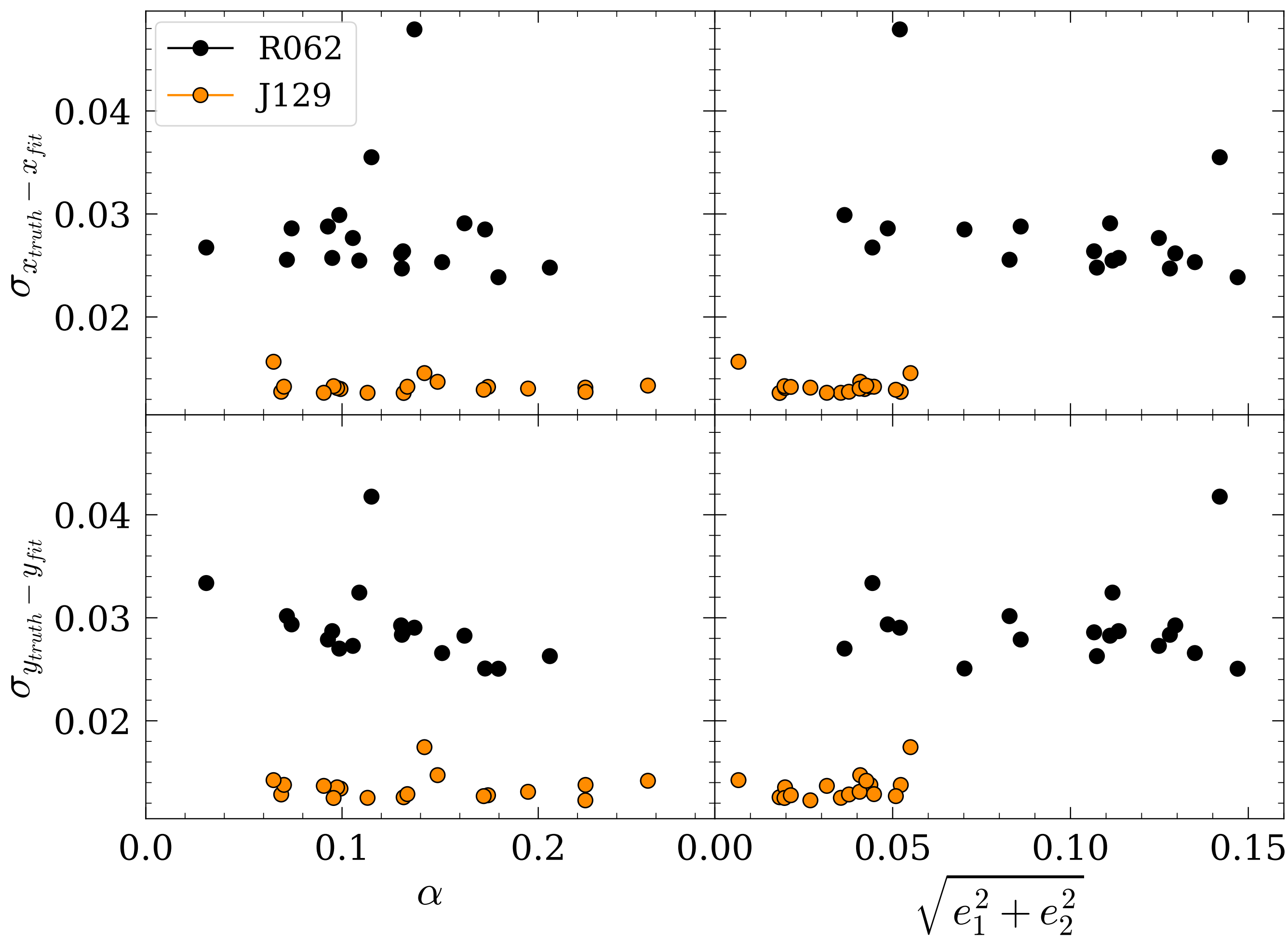

**Figure 12.** Pixel offsets for each SCA in Figure 7 compared to the corresponding ePSF asymmetry ($\alpha$) and ellipticity ($\sqrt{e_1^2 + e_2^2}$). The R062 band has a more asymmetric PSF than the J129 band (Figure 8; Appendix A, Tables 4, 5).

These results emphasize the importance of considering the PSF in terms focal plane location, in addition to how it changes between two locations. Further, certain filters will be better for accurately and precisely determining object locations than others. Future work with Roman images that involves developing ePSFs, whether they are from the OpenUniverse2024 simulations or real images, should take this effect into account. In other words, stars that are used for generating

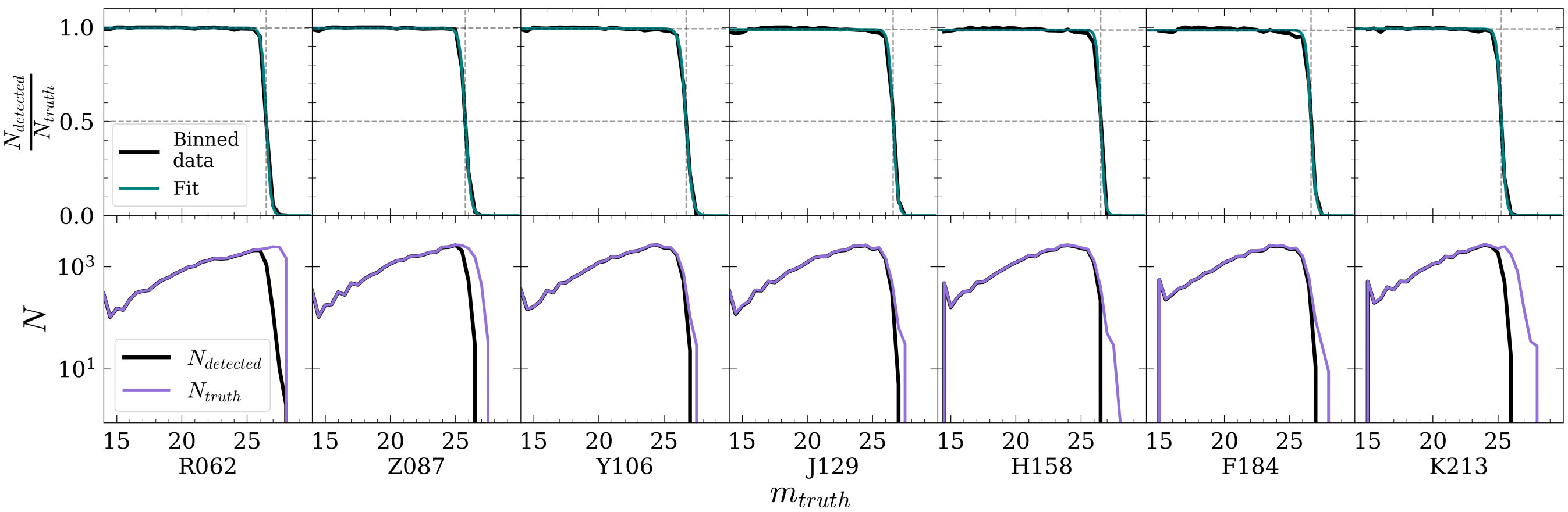

**Figure 13.** Top: fraction of all stars detected binned every 0.5 mag (black) and Equation (3) fit to this line (teal; Table 3). The horizontal dashed line near $\frac{N_{\mathrm{detected}}}{N_{\mathrm{truth}}} \approx 1$ is $\eta_0$, and the vertical dashed line that intersects with the line at $\frac{N_{\mathrm{detected}}}{N_{\mathrm{truth}}} = 0.5$ is $m_{50}$. Bottom: the number of stars detected (black) and the total number of stars in the truth catalog (purple) as a function of truth catalog magnitude. Data are binned every 0.5 mag.

**Table 3**
Parameters from the Fit to Our Binned Detection Efficiency Curves Using Equation (3)

| Band | $\eta_0$ | $m_{50}$ | $\tau$ |
|---|---|---|---|
| R062 | 0.9966 | 26.48 | 0.1691 |
| Z087 | 0.9961 | 25.76 | 0.2004 |
| Y106 | 0.9930 | 26.70 | 0.2262 |
| J129 | 0.9883 | 26.59 | 0.1729 |
| H158 | 0.9858 | 26.52 | 0.1453 |
| F184 | 0.9851 | 26.65 | 0.1793 |
| K213 | 0.9915 | 25.26 | 0.1714 |

**Note.** All fit parameter errors are of order $10^{-4}$ or smaller.

the ePSF should be chosen in part based on their location on the focal plane.

This work suggests that observations associated with ePSF generation should be designed specifically for this purpose. Candidate stars for ePSF generation should be chosen in advance and placed at specific detector locations in accordance with the expected PSF FWHM (Figure 1), asymmetry (Figure 9), and ellipticity pattern (Section 3.3.3). Ideal locations for these would be near inflection areas, for example, where $e_1$ and $e_2$ (Section 3.3.3) change sign or where the FWHM changes quickly. We expect that "anchoring" PSFs at key locations will allow interpolation between PSFs that is precise enough for most applications. If this is not possible, a sufficiently fine grid should produce similar precision in results.

Roman's photometric touchstone fields will be used to generate the official ePSF library. These fields will require >100 stars per SCA between $16 < m_{\mathrm{AB}} < 22$ (S. Casertano et al. 2021), with aggregate $\mathrm{S/N} > 300$. These fields will be observed regularly, and observations will be dithered to characterize subpixel response. It will be useful to repeat the tests carried out in this work during the calibration phase of the mission in order to determine the optimal methodology for generating the on-orbit ePSF library.

It is important to note that although the SOC will provide an ePSF function for community use, if empirically generating a new ePSF directly from HLTDS images is preferred for a particular analysis (e.g., if the Roman PSF changes significantly over the mission's duration), it will not be possible to achieve Roman's expected precision by straightforwardly using the stars in one image. The spatial variation means that if this method is chosen, subdivision of the image must occur in order to develop robust ePSFs; however, if the image is subdivided and the magnitude range of stars is not increased, then there are not enough stars in each region of a single HLTDS image to generate a precise ePSF.

## 4. Conclusions and Future Work

In this work, we develop a library of ePSFs compatible with the OpenUniverse2024 HLTDS Roman simulations and show that it is possible to achieve the expected <1% flux recovery (∼millimagnitude-level) precision using these ePSFs for photometry, using fractional stellar fluxes when sky noise is subdominant. We empirically demonstrate the importance of considering the spatial variation of Roman's PSF in analyses by dividing the images into grids of different sizes (Table 1) and generating PSFs for progressively smaller areas on the SCAs. Of the ePSFs generated in this work, the $8 \times 8$ grid performs the best; the scatter in flux recovery decreases by up to 20% between the $1 \times 1$ grid and the $8 \times 8$ grid (Figure 4, Table 2).

We successfully recover the effects associated with spatial variation of the injected PSF, including asymmetry and ellipticity, in the OpenUniverse2024 simulations on flux measurements. These effects include both accuracy and precision of object coordinate recovery (Figures 4, 7, and 9). We quantify the effects of nonlinearity due to photometry and recover fluxes to within a slope $|s_{\mathrm{NL}}| < 1.93 \times 10^{-3}\,\mathrm{dex}^{-1}$ in all bands (Figure 2). Given that the count-rate nonlinearity requirement for Roman is less than 0.3% over 11 mag, this value indicates that further work is still needed to reduce or characterize any nonlinear effects due to photometric methodology. We also find a possible color dependence that is strongest in R062 (Figure 6). In R062, we find $s_{\mathrm{color}} = -4.73 \times 10^{-2}$, while $\hat{\sigma}$ in the same measured region is $\hat{\sigma} = 1.22 \times 10^{-2}$. This suggests that a color-dependent PSF may be necessary. Finally, we also show that Source Extractor (E. Bertin & S. Arnouts 1996) detects nearly 100% of stars down to approximately 26th magnitude, after which point very few are successfully detected (Figure 13).

We emphasize that this work is not intended to be an exhaustive investigation of the Roman PSF but rather serve as an initial "benchmark" analysis. It is intended to be a first pass to obtain baselines and describe potential difficulties with using it. Forthcoming work on the Roman PSF will address the deficiencies in this work. For example, we do not consider dithered observations as a part of this work, even though increased uncertainty in stellar position is associated with the lack of dithered observations (J. Anderson & I. R. King 2000). The final HLTDS observing strategy will include dithering by rotation, and it will be possible to use these repeated measurements of the same stars to break the degeneracy between position and PSF shape. It is promising that even with this simpler methodology, we achieve the expected flux precision from Roman, and we look forward to results from future work refining Roman photometry.

## Acknowledgments

L.A. thanks Jay Anderson for the helpful discussion about ePSFs and Mike Jarvis for discussions about GalSim PSFs. L. A. also thanks collaborators Rick Kessler, Stefano Casertano, and Cole Meldorf for the helpful comments and discussions. The authors thank the reviewer for their useful comments.

This work is supported by NASA under award number 80GSFC24M0006. Additionally, funding for the Roman Supernova Project Infrastructure Team has been provided by NASA under contract to 80NSSC24M0023. A.K. and M.T. were funded by NASA under JPL Contract Task 70–711320, "Maximizing Science Exploitation of Simulated Cosmological Survey Data Across Surveys."

This research used resources of the National Energy Research Scientific Computing Center, which is supported by the Office of Science of the U.S. Department of Energy using award number HEP-ERCAP32751.

*Facility:* NERSC.

*Software:* astropy (Astropy Collaboration et al. 2013, 2018, 2022), matplotlib (J. D. Hunter 2007), numpy (C. R. Harris et al. 2020), pandas (W. McKinney 2010; The pandas development team 2020), photutils (L. Bradley et al. 2024), scipy (P. Virtanen et al. 2020), Source extractor (E. Bertin & S. Arnouts 1996).

## Appendix A
## Asymmetry Indices

In this work, we quantify asymmetry by dividing a PSF image into a $2 \times 2$ grid, summing the pixels in each quadrant and comparing unique combinations of those sums. We define this mathematically in Equation (2). The values of $\alpha$ computed at the center of each SCA (2044, 2044) in each band are shown in Tables 4 and 5.

**Table 4**
Asymmetry Indices for the Injected GalSim PSFs at the Centers of All SCAs (2044, 2044)

| SCA | R062 | Z087 | Y106 | J129 | H158 | F184 | K213 |
|---|---|---|---|---|---|---|---|
| 1 | 0.1372 | 0.1027 | 0.0861 | 0.0725 | 0.0621 | 0.0606 | 0.0515 |
| 2 | 0.1111 | 0.0828 | 0.0701 | 0.0581 | 0.0476 | 0.0492 | 0.0454 |
| 3 | 0.1135 | 0.0850 | 0.0729 | 0.0611 | 0.0519 | 0.0513 | 0.0437 |
| 4 | 0.1115 | 0.0821 | 0.0682 | 0.0597 | 0.0499 | 0.0528 | 0.0472 |
| 5 | 0.0847 | 0.0630 | 0.0546 | 0.0428 | 0.0351 | 0.0384 | 0.0387 |
| 6 | 0.1090 | 0.0831 | 0.0720 | 0.0577 | 0.0511 | 0.0414 | 0.0350 |
| 7 | 0.0846 | 0.0645 | 0.0609 | 0.0516 | 0.0438 | 0.0440 | 0.0426 |
| 8 | 0.0705 | 0.0551 | 0.0441 | 0.0401 | 0.0303 | 0.0366 | 0.0313 |
| 9 | 0.0897 | 0.0639 | 0.0500 | 0.0420 | 0.0337 | 0.0276 | 0.0248 |
| 10 | 0.1770 | 0.1379 | 0.1152 | 0.0992 | 0.0846 | 0.0853 | 0.0758 |
| 11 | 0.1447 | 0.1094 | 0.0921 | 0.0765 | 0.0631 | 0.0646 | 0.0591 |
| 12 | 0.1200 | 0.0932 | 0.0793 | 0.0629 | 0.0546 | 0.0519 | 0.0466 |
| 13 | 0.2202 | 0.1763 | 0.1498 | 0.1280 | 0.1052 | 0.1117 | 0.1001 |
| 14 | 0.1842 | 0.1441 | 0.1205 | 0.1044 | 0.0900 | 0.0893 | 0.0777 |
| 15 | 0.1331 | 0.1027 | 0.0858 | 0.0745 | 0.0609 | 0.058 | 0.0537 |
| 16 | 0.2487 | 0.1980 | 0.1728 | 0.1492 | 0.1290 | 0.1290 | 0.1120 |
| 17 | 0.2178 | 0.1757 | 0.1495 | 0.1299 | 0.1131 | 0.1101 | 0.0957 |
| 18 | 0.1591 | 0.1276 | 0.1105 | 0.0964 | 0.0852 | 0.0739 | 0.0672 |

**Note.** PSFs are oversampled by a factor of 3 and convolved with the Roman pixel response.

**Table 5**
Asymmetry Indices for the ePSFs Generated in This Work, Centered at (1788.5 px, 1788.5 px), of the 8 × 8 Grid

| SCA | R062 | Z087 | Y106 | J129 | H158 | F184 | K213 |
|---|---|---|---|---|---|---|---|
| 1 | 0.0986 | 0.1219 | 0.134 | 0.1314 | 0.1127 | 0.1020 | 0.0822 |
| 2 | 0.0950 | 0.0977 | 0.1020 | 0.0992 | 0.0872 | 0.0840 | 0.0710 |
| 3 | 0.1088 | 0.0929 | 0.1154 | 0.1130 | 0.1032 | 0.0915 | 0.0765 |
| 4 | 0.0743 | 0.0928 | 0.1011 | 0.0975 | 0.0879 | 0.0841 | 0.0740 |
| 5 | 0.1055 | 0.0860 | 0.0773 | 0.0690 | 0.0583 | 0.0527 | 0.0461 |
| 6 | 0.0719 | 0.0832 | 0.0661 | 0.0957 | 0.0820 | 0.0788 | 0.0373 |
| 7 | 0.0928 | 0.0878 | 0.0947 | 0.0907 | 0.0783 | 0.0710 | 0.0625 |
| 8 | 0.1150 | 0.0878 | 0.0745 | 0.0704 | 0.0524 | 0.0544 | 0.0476 |
| 9 | 0.0308 | 0.0526 | 0.0591 | 0.0651 | 0.0552 | 0.0496 | 0.0372 |
| 10 | 0.1369 | 0.1701 | 0.1749 | 0.1744 | 0.1563 | 0.1368 | 0.1263 |
| 11 | 0.1311 | 0.1470 | 0.1533 | 0.1487 | 0.1238 | 0.1156 | 0.1638 |
| 12 | 0.1305 | 0.1231 | 0.1335 | 0.1333 | 0.1223 | 0.1040 | 0.0922 |
| 13 | 0.1729 | 0.2131 | 0.2266 | 0.2240 | 0.1959 | 0.1777 | 0.1589 |
| 14 | 0.1624 | 0.1937 | 0.1941 | 0.1948 | 0.1648 | 0.1480 | 0.1227 |
| 15 | 0.1301 | 0.1446 | 0.1540 | 0.1420 | 0.1257 | 0.1127 | 0.0962 |
| 16 | 0.2059 | 0.2379 | 0.2514 | 0.2559 | 0.2294 | 0.1984 | 0.1888 |
| 17 | 0.1797 | 0.2132 | 0.2281 | 0.2241 | 0.1981 | 0.1778 | 0.1616 |
| 18 | 0.1510 | 0.1769 | 0.1783 | 0.1722 | 0.1555 | 0.1243 | 0.1085 |

## Appendix B
## Ellipticity

The ellipticity $e$, in terms of the ratio of the semimajor ($a$) and semiminor ($b$) axes, is defined as

$$e = e_1 + \mathrm{i}e_2 = \frac{1 - q^2}{1 + q^2}\exp(2\mathrm{i}\phi), \tag{B1}$$

where $q = b/a$ and $\phi$ is the angle of the major axis with respect to the $x$-axis. In practice, $e_1$ and $e_2$ are measured as

$$e_1 = \frac{M_{xx} - M_{yy}}{M_{xx} + M_{yy}} \tag{B2}$$

$$e_2 = \frac{2M_{xy}}{M_{xx} + M_{yy}}, \tag{B3}$$

where $M_{xx}$, $M_{yy}$ and $M_{xy}$ are adaptive Gaussian moments as measured by GalSim, where

$$M_{xy} = \int\int dxdy\ xy\ w(x, y)P(x, y) \tag{B4}$$

$$M_{xx} = \int\int dxdy\ x^2\ w(x, y)P(x, y), \tag{B5}$$

and similarly for $M_{yy}$. The origin is chosen to be where the first-order moments vanish. The values of $e$ computed for some of the ePSFs generated in this work are shown in Table 6.

**Table 6**
Ellipticity Values for the ePSFs Generated in This Work, Centered at (1788.5 px, 1788.5 px), of the 8 × 8 Grid

| SCA | R062 | | Z087 | | Y106 | | J129 | | H158 | | F184 | | K213 | |
|---|---|---|---|---|---|---|---|---|---|---|---|---|---|---|
| | $e_1$ | $e_2$ | $e_1$ | $e_2$ | $e_1$ | $e_2$ | $e_1$ | $e_2$ | $e_1$ | $e_2$ | $e_1$ | $e_2$ | $e_1$ | $e_2$ |
| 1 | −0.0331 | −0.0152 | −0.0268 | −0.0125 | −0.0230 | −0.0083 | −0.0178 | −0.0037 | −0.0161 | −0.0046 | −0.0151 | 0.0005 | −0.0144 | 0.0018 |
| 2 | −0.0630 | −0.0944 | −0.0521 | −0.0636 | −0.0400 | −0.0429 | −0.0317 | −0.0277 | −0.0239 | −0.0145 | −0.0268 | −0.0118 | −0.0248 | −0.0041 |
| 3 | −0.0004 | −0.1118 | −0.0015 | −0.0779 | −0.0101 | −0.0560 | −0.0109 | −0.0337 | −0.0222 | −0.0149 | −0.0279 | −0.0090 | −0.0347 | −0.0019 |
| 4 | −0.0220 | −0.0433 | −0.0236 | −0.0298 | −0.0163 | −0.0214 | −0.0149 | −0.0129 | −0.0100 | −0.0065 | −0.0102 | −0.0068 | −0.0098 | −0.0008 |
| 5 | −0.0483 | −0.1151 | −0.0437 | −0.0770 | −0.0314 | −0.0523 | −0.0203 | −0.0318 | −0.0185 | −0.0161 | −0.0224 | −0.0099 | −0.0206 | −0.0010 |
| 6 | 0.0310 | −0.0768 | 0.0080 | −0.0571 | 0.0017 | −0.0339 | −0.0037 | −0.0192 | −0.0124 | 0.0010 | −0.0205 | 0.0025 | −0.0250 | 0.0080 |
| 7 | −0.0247 | −0.0824 | −0.0192 | −0.0559 | −0.0171 | −0.0428 | −0.0104 | −0.0297 | −0.0025 | −0.0183 | −0.0072 | −0.0162 | −0.0047 | −0.0089 |
| 8 | −0.0555 | −0.1307 | −0.0438 | −0.0837 | −0.0332 | −0.0645 | −0.0189 | −0.0394 | −0.0173 | −0.0218 | −0.0137 | −0.0159 | −0.0141 | −0.0037 |
| 9 | 0.0135 | −0.0422 | 0.0097 | −0.0253 | 0.0013 | −0.0186 | −0.0043 | −0.0051 | −0.0080 | 0.0041 | −0.0112 | 0.0032 | −0.0135 | 0.0130 |
| 10 | −0.0494 | 0.0164 | −0.0372 | 0.0162 | −0.0269 | 0.0133 | −0.0196 | 0.0085 | −0.0168 | 0.0064 | −0.0178 | 0.0030 | −0.0153 | 0.0018 |
| 11 | −0.0851 | −0.0643 | −0.0611 | −0.0435 | −0.0475 | −0.0323 | −0.0347 | −0.0216 | −0.0280 | −0.0187 | −0.0293 | −0.0092 | −0.0325 | −0.0104 |
| 12 | −0.0299 | −0.1244 | −0.0231 | −0.0852 | −0.0195 | −0.0627 | −0.0219 | −0.039 | −0.0247 | −0.0259 | −0.0315 | −0.0202 | −0.0370 | −0.0133 |
| 13 | −0.0522 | 0.0469 | −0.0358 | 0.0361 | −0.0273 | 0.0319 | −0.0163 | 0.0213 | −0.0134 | 0.0130 | −0.0115 | 0.0121 | −0.0087 | 0.0050 |
| 14 | −0.1104 | −0.0128 | −0.0860 | −0.0035 | −0.0546 | −0.0035 | −0.0401 | −0.0072 | −0.0276 | −0.0074 | −0.0264 | −0.0071 | −0.0251 | −0.0105 |
| 15 | −0.0662 | −0.1112 | −0.0534 | −0.0838 | −0.0421 | −0.0617 | −0.0320 | −0.0448 | −0.0306 | −0.0376 | −0.0287 | −0.0243 | −0.0303 | −0.0264 |
| 16 | −0.0664 | 0.0845 | −0.0431 | 0.0674 | −0.0262 | 0.0518 | −0.0204 | 0.0374 | −0.0074 | 0.0232 | −0.0030 | 0.0180 | −0.0073 | 0.0124 |
| 17 | −0.1380 | 0.0506 | −0.1006 | 0.0444 | −0.0713 | 0.0336 | −0.0492 | 0.0178 | −0.0298 | 0.0063 | −0.0244 | 0.0031 | −0.0160 | −0.0012 |
| 18 | −0.1218 | −0.0582 | −0.0872 | −0.0449 | −0.063 | −0.0352 | −0.0392 | −0.0324 | −0.0307 | −0.0273 | −0.0233 | −0.0181 | −0.0203 | −0.0222 |

## ORCID iDs

L. Aldoroty https://orcid.org/0000-0003-0183-451X
D. Scolnic https://orcid.org/0000-0002-4934-5849
A. Kannawadi https://orcid.org/0000-0001-8783-6529
R. A. Knop https://orcid.org/0000-0002-3803-1641
B. M. Rose https://orcid.org/0000-0002-1873-8973
R. Hounsell https://orcid.org/0000-0002-0476-4206
M. A. Troxel https://orcid.org/0000-0002-5622-5212